\documentclass[conference]{HPCSTrans}
\IEEEoverridecommandlockouts
\usepackage{amsmath,amssymb,enumerate}
\usepackage{cite,graphics,graphicx}

\usepackage[utf8]{inputenc}
\usepackage[T1]{fontenc}
\usepackage{tabulary}
\usepackage{pgfplots}
\usepackage{tikz}
\usetikzlibrary{arrows, positioning, calc}
\usepackage{cite}
\ifCLASSINFOpdf
  \graphicspath{{images/}}
\else
\fi

\usepackage{stfloats}
\usepackage{url}

\begin{document}
%
\title{Parallel decoder for Low Density Parity Check Codes: A MPSoC study}

\author{
\IEEEauthorblockN{Sudeep Kanur\IEEEauthorrefmark{1}, 
Georgios Georgakarakos\IEEEauthorrefmark{2}, 
Antti Siirilä \IEEEauthorrefmark{3},
Jérémie Lagravière \IEEEauthorrefmark{4}, 
\\
Kristian Nybom \IEEEauthorrefmark{5},
Sébastien Lafond \IEEEauthorrefmark{6} and
Johan Lilius \IEEEauthorrefmark{7}}
\IEEEauthorblockA{\IEEEauthorrefmark{1},\IEEEauthorrefmark{5},\IEEEauthorrefmark{6},\IEEEauthorrefmark{7} \textit{\AA bo Akademi University, Turku, Finland.} \\
Email: \textit{firstname.lastname@abo.fi}}
\IEEEauthorblockA{\IEEEauthorrefmark{2},\IEEEauthorrefmark{3},\IEEEauthorrefmark{4}, \textit{Turun Yliopisto, Turku, Finland.} \\
Email: \textit{firstname.lastname@utu.fi}}
}


%


\maketitle

\begin{abstract}
The near channel performance of Low Density Parity Check Codes (LDPC) has motivated its wide applications. Iterative decoding of LDPC codes provides significant implementation challenges as the complexity grows with the code size. Recent trends in integrating Multiprocessor System on Chip (MPSoC) with Network on Chip (NoC) gives a modular platform for parallel implementation. This paper presents an implementation platform for decoding LDPC codes based on HeMPS, an open source MPSoC framework based on NoC communication fabric. Reduced minimum sum algorithm is used for decoding LDPC codes and simulations are performed using HeMPS tool. The data rate and speedup factor measured for decoding a rate 1/2 LDPC code characterised by $252 \times 504$ parity matrix is presented.
\end{abstract}
\vspace{0.1in}
\begin{keywords}
Low Density Parity Check codes (LDPC), Minimum sum decoding algorithm, HeMPS, MPSoC, NoC, Message passing interface (MPI)
\end{keywords}

%
\IEEEpeerreviewmaketitle

\section{INTRODUCTION}

Reliable transmission of data over a noisy communication channel is one of the central goals a communication system strives to achieve. Forward Error Correction schemes are used in broadcasting and communication systems to increase the bandwidth and improve the efficiency. Low Density Parity Check (LDPC) codes, introduced by Gallager \cite{Gallager1962}, are forward error correction codes that have been proved to achieve performance close to Shannon's limit \cite{Mackay1997}. This performance has motivated its use in application areas ranging from long range satellite transmission to terrestrial broadcasting \cite{Andrews2007}. Second generation Digital Video Broadcasting (DVB) standards has used LDPC for satellite, cable, mobile and terrestrial broadcasting channels.

LDPC codes are linear block codes and have the property that the error correction performance increases with increasing code lengths. Moreover, if special care is taken when designing the codes, encoding can be made trivial. However, decoding LDPC codes is a
NP-complete problem \cite{Berlekamp1978}. Iterative decoding schemes
are used in decoding, but they bring significant implementation
challenges. On the positive side, iterative decoding schemes
provide modularity for parallel implementation which many
implementation platforms have exploited for faster throughput. 

The work in \cite{Dielissen2006} and \cite{Kienle2005} presents IP cores for second generation Digital Video Broadcasting for satellite (DVB-S2) standards on an ASIC, while \cite{Gomes2007} present the same on flexible platforms such as field programmable gate arrays
(FPGAs). Software Defined Radio (SDR) and GPU implementations have also been created for the standard \cite{Gronroos2011}. While afore mentioned works have achieved real time throughput rates, they come at the cost of increased design time as the IP cores are hand-coded in the cases of FPGA or increased hardware size as in the case of GPUs and SDR.

With the advancements in semiconductors and CAD design, Multiprocessor System on Chip (MPSoC) with underlying Network on Chip (NoC) platform is gaining popularity in embedded systems. The platform provides modularity at processor level and can be exploited for the implementation of the iterative decoding of LDPC. HeMPS is an open source framework targeting MPSoC customisation that includes the platform comprising of NoC, processors and DMA, the embedded software comprising of microkernel and applications and a dedicated CAD tool to generate required binaries for implementation on FPGA platforms and perform debugging\cite{Wachter2011}. Communication between processors is done using the NoC communication structure and is implemented using message passing routines. In addition the platform supports static and dynamic task mapping and C or SystemC simulation models for processors and memories \cite{Carara2009}. 

Exploitation of the HeMPS modularity coupled with decreased time and the cost of implementation is the main goal of this paper. The decoding of a rate 1/2 LDPC code, defined by a $252 \times 504$ the parity check matrix with progressive edge growth construction, is used for the study. The work simulates sequential and parallel implementation of the algorithm and measures throughput data rates. In addition, a scale-up factor is presented and data rates are compared with a message passing interface library (MPI) implementation on a desktop machine to get a perspective. The paper is organised as follows. Section II provides details on the decoding algorithm. Section III introduces the MPSoC framework. Section IV presents the experimental setup and details mapping of the algorithm on the framework. Section V presents observations and results while section VI concludes the paper.

\section{LDPC CODES}

\begin{figure}[tb]
\centering
\includegraphics[width=3in]{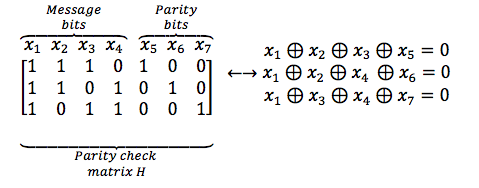}
\caption{Parity Check Matrix, $H$}
\label{matrix}
\end{figure}

\begin{figure}[tb]
\centering
\includegraphics[width=3in]{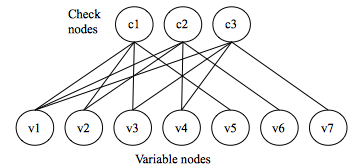}
\caption{Tanner graph representation of parity matrix $H$}
\label{tannergraph}
\end{figure}

A LDPC code of length $n$ bits consists of $k$ bits of information and $n-k$ bits of redundancy called parity bits. The code rate $R = k/n$ gives the fraction of useful information among the transmitted information. The relationship between information and parity bits is given by linear equations and these linear equations can be represented in a matrix form, known as parity check matrix $H$. Fig.\ref{matrix} illustrates the parity-check matrix H for a Hamming code with code length $n=7$ and information length $k=4$. For a valid code word $x$, the linear equations shown in Fig. \ref{matrix} are satisfied. The validity of the codeword can be also checked using matrix relation $H\cdot x^{T} = 0$.
Unlike the Hamming code, the parity matrix of LDPC codes is constructed using $m \times n$ sparse matrix. A sparse matrix has a large number of  zero elements compared with non-zero elements and the number of non-zero entries in $H$ grow linearly as $O(n)$. Tanner graphs can be constructed from parity check matrix $H$ by connecting \emph{check nodes}, given by the row indices of $H$ with \emph{variable nodes}, given by the column indices of $H$ and is drawn according to the following rule: a check node $i$ is connected to variable node $j$ whenever the element $h_{ij}$ of $H$ is 1, where $i$ and $j$ represent rows and columns of the parity matrix. The Tanner graph shown in Fig. \ref{tannergraph}, represent the entire LDPC code and can help in understanding decoder algorithms for LDPC codes.

Gallager proposed an iterative probabilistic decoding scheme based on message or belief propagation to decode LDPC codes\cite{Gallager1962}. Several decoding techniques have been proposed since then and all of the effective decoding strategies for LDPC are message passing algorithms. The Minimum Sum algorithm has been proved to provide good performance with the usage of minimal hardware\cite{Anastasopoulos2001}. The reduced scan minimum sum algorithm, a variation of the minimum sum algorithm, was proposed for systems with memory constraints \cite{Xhuang2006}. This algorithm is a rearrangement of the equations of the minimum sum algorithm and has less memory footprint than the original algorithm. The reduced footprint comes at the cost of reduced performance, and in implementations constrained with memory size, this algorithm can be used. 

The reduced minimum sum algorithm (RMSA) works by passing messages or beliefs between the check nodes and variable nodes of the tanner graph. RMSA operates in the logarithm domain and messages passed between nodes known as Log Likelihood Ratios (LLRs) which are defined as,
\begin{equation}\label{LLR}
\Lambda(x_{i}) = log(\frac{P(x_{i}=0|y_{i})}{P(x_{i}=1|y_{i})}), \hspace{10mm} -\infty \leq \Lambda \leq \infty
\end{equation}
where $i = 1 \ldots n$, $x$ is the code word, $y$ is the channel output and $P(a|b)$ is the conditional probability function. For a decoded message $\hat{x}$, LLRs exhibit the property that as $\Lambda  \rightarrow \infty$, $P(\hat{x} = 0) \rightarrow 1$ and $\Lambda  \rightarrow -\infty$, $P(\hat{x} = 1) \rightarrow 1$. For $\Lambda  = 0$, $\hat{x}$ could be either 0 or 1.

Let us consider the Tanner graph shown in Fig.\ref{tannergraph}. Let $C(v)$ denotes the set of check nodes which connect to variable nodes $v$. Similarly let $V(c)$ denote the set of variable nodes that connect to check nodes $c$. Let $C(v)\setminus c$ represent all members of set $C(v)$ except $c$, while $V(c)\setminus v$ represent all members of set $V(c)$ expect $v$. The RMSA decoding algorithm for $j$ iterations can be summarised in following steps.
\begin{enumerate}
\item \emph{Initialisation}: Each variable node $v$ is assigned an \emph{a priori} LLR according to
$$\Lambda_{v}^{(0)} = \Lambda^{(0)}$$ 
Similarly check node messages are initialised to zero, 
$$ \Lambda_{c \rightarrow v}^{(0)} = 0$$ 

\item \emph{Check node update}: For each check node $c$ and for each $v \in V(c)$, compute 
\begin{equation}
\begin{split}
\Lambda_{c \rightarrow v}^{(j)}  = (\prod_{v' \in V(c)\setminus v} sign(\Lambda_{v}^{(j-1)}  - \Lambda_{c \rightarrow v'}^{(j-1)}) \\
\times \underset{v' \in V(c)\setminus v}{min}|\Lambda_{v}^{(j-1)}  - \Lambda_{c \rightarrow v'}^{(j-1)}|
\end{split}
\end{equation}

\item \emph{Variable node update}: For each variable node $v$, compute 
\begin{equation}
\Lambda_{v}^{(j)}  = \Lambda^{(0)}  + \sum_{c \in C(v)}\Lambda_{c \rightarrow v}^{(j)}  
\end{equation}

\item \emph{Decision}: Obtain intermediate $x_{i}$ such that if $
\Lambda^{j}_{v} \geq 0$, then $\hat{x_{i}} = 0$. Otherwise $\hat{x_{i}} = 1$. Check for the condition $H\cdot \hat{x}^{T} = 0$. If the condition is satisfied $\hat{x}$ is a valid codeword, else go to step 2 and iterate until iteration limit is reached or until a valid codeword is found. 
\end{enumerate}

A careful observation of the algorithm at step 2 reveals that the check node computation of a given check node is independent of others in a given iteration. Hence, in an ideal case, all the check node computations can run in parallel in a given iteration with messages passing between check nodes and variable nodes.

\section{HEMPS FRAMEWORK}

\begin{figure}[tb]
\centering
\includegraphics[width=3in]{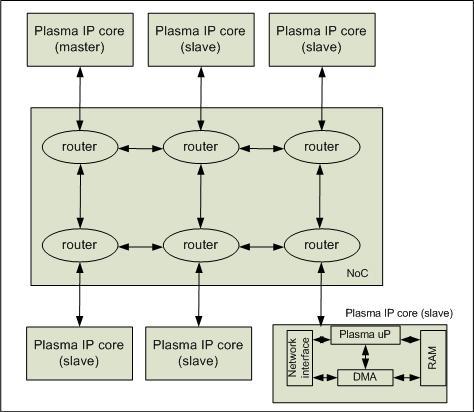}
\caption{HeMPS NOC architecture}
\label{plasma}
\end{figure}

The HeMPS framework provides a MPSoC framework with the network on chip like connectivity. The HERMES NoC architecture  is used to connect “Plasma” (MIPS-like) processors in a 2D mesh topology as shown in Fig. \ref{plasma}. The processing elements (PEs) are named Plasma-IP. Each PE in the network contains a plasma processor, a local memory, a DMA controller, and a network interface. The DMA controller is used for fast data transfer of packets between local memory and the PE’s network interface. Among the PEs one ‘master’ (PlasmaIP-MS) is responsible for managing resources while the rest ‘slaves’ (PlasmaIP-SL) are executing the applications.

During operation, the master reads the distributed application source (split in several tasks) from an external task repository, and allocates tasks to the slaves. Each slave runs a microkernel, which supports multitasking and task communication. The master also runs a microkernel, but does not execute applications tasks. The microkernel segments the memory in pages, which is allocated for itself (first page) and several tasks (subsequent pages). Each Plasma-IP has a task table, with the location of local and remote tasks. The microkernel protects memory pages. All communication between tasks is handled through a custom blocking message passing interface. The messages exchanged between processors are constrained to 128 bytes each. The kernel is described mostly in C and some special functions such as interruption treatment and context saving are described in assembly.

\section{IMPLEMENTATION}

 \subsection{Mapping}
 For a parallel scalable implementation of decoding algorithm, the option of assigning each node to a separate processing element is clearly not attractive. Such an implementation introduces high communication overhead between the processing elements. Instead, the approach proposed is to group several check nodes together and to execute each group on separate processing elements in a homogenous MPSOC architecture.The goal is to have scalability and stable performance enhancements in LDPC decoding. In principle, the mapping involves dividing the decoding algorithm between one master processor and several slave processors. The master processor is in charge of scattering and gathering of data and computation of the steps 1, 3 and 4 of the algorithm while slave processors are in charge of unpacking the data and processing the check nodes. The arrangement resembles a star-based network configuration between master and slave processors.

\begin{figure}[tb]
\centering
\includegraphics[width=3in]{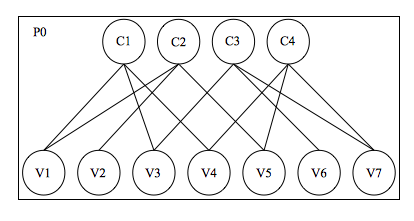}
\caption{Sequential programming model for LDPC decoder}
\label{sequential}
\end{figure}

\begin{figure}[tb]
\centering
\includegraphics[width=3in]{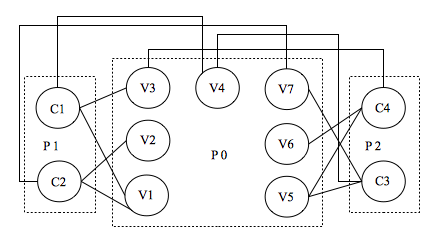}
\caption{Parallel programming model for LDPC decoder}
\label{parallel2}
\end{figure}

\begin{figure}[tb]
\centering
\includegraphics[width=3.5in]{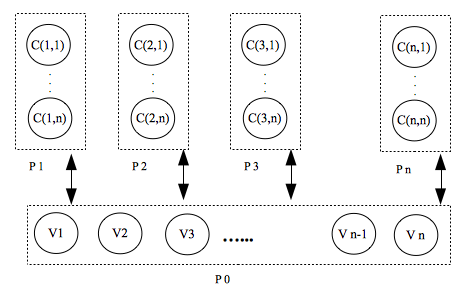}
\caption{Generalised parallel programming model for LDPC decoder}
\label{parallel}
\end{figure}

The check node grouping decision is made by dividing the number of check nodes by the number of slave processors. Since the processing follows a homogeneous concept, each slave processor is assigned the same number of check nodes. Consider the Tanner graph shown in Fig.\ref{sequential}. For the sake of convenience, a 4 by 7 parity check matrix is chosen. Fig. \ref{parallel2} gives an example of an implementation of the algorithm executed on three processing elements (\textit{P0, P1 and P2}). \textit{P0} executes all the variable nodes, and communicates with all the check nodes and forms the central process. \textit{P1} and \textit{P2} are in charge of the execution of two check nodes each. Fig.\ref{parallel} shows the same for a generalised case.

 \subsection{Simulation Setup}
 The distributed LDPC decoder application is mapped as described in the
previous section and transferred to HeMPS simulation platform for the
test and evaluation. For the experimentation, a parity check matrix of
size 252 by 504 was chosen \cite{PegGirReg}. LDPC codes, being linear
codes gives better performance as the size of the code increases.
However, the simulator platform restricts the memory size, and due to
this restriction this parity matrix was chosen. The restriction on
memory size also motivated the use of RMSA algorithm instead of the minimum sum algorithm for decoding LDPC codes.
To get a perspective of HeMPS platform, a measurement was carried out
on a desktop machine using MPI as the framework. This gave an idea of
overheads that HeMPS platform puts on the blocking message passing interface. The desktop machine has the following property: two CPUs \textit{Intel(R) Xeon(R) CPU E5-2620 0 @ 2.00GHz}, in total it represents 12 physical cores, and 24 cores (hyperthreading),
running \textit{Linux Ubuntu 12.04 - Kernel 3.2.0-38-generic-pae} operating system.

HeMPS can perform a System-C based simulation for a given MPSoC
configuration. HeMPS API allows various mapping options in a
configurable NoC based MPSoC environment. Since the mapping approach proposed features a
master kernel communicating with several slave kernels, the master
kernel is placed at the center of the processing elements array, while
the slave kernels are allocated in processing elements (PE) around the
central PE. The PEs are configured as Plasma IP cores at a clock speed
of 100 MHz and a maximum memory page size of 16KB (program and data).
Each LDPC decoder kernel uses 1 memory page size per core. The input
of the decoder is the LLR values of a single codeword from the channel
receiver. These LLR values are hard coded into the memory page of the master node for the sake of simplicity. Using a system call that measures clock cycles, a performance
evaluation of several check node grouping options can be obtained. The
measurements performed here are the worst case measurements and are    
calculated for the maximum iterations of 30. The measurements are taken for increasing number of slave processors until the performance starts to degrade. 

The speed-up of each grouping option is measured with respect to the normal (sequential) LDPC decoding application. The sequential application features the LDPC decoder mapped on a single Plasma IP core.

%

Seven different scenarios have been considered for measurement. All the scenarios run 1 task per PE only. As the number of check nodes needs to be divisible by the processors, certain processor configurations could not be measured. The scenarios considered were:

\begin{enumerate}
	\item sequential LDPC decoding application: 1 task, single PE
	\item 3-PE allocation (1 master-2 slaves)
	\item 4-PE allocation (1 master-3 slaves)
	\item 5-PE allocation (1 master-4 slaves)
	\item 7-PE allocation (1 master-6 slaves)
	\item 8-PE allocation (1 master-7 slaves)
	\item 10-PE allocation (1 master-9 slaves)
\end{enumerate}

\section{RESULTS}
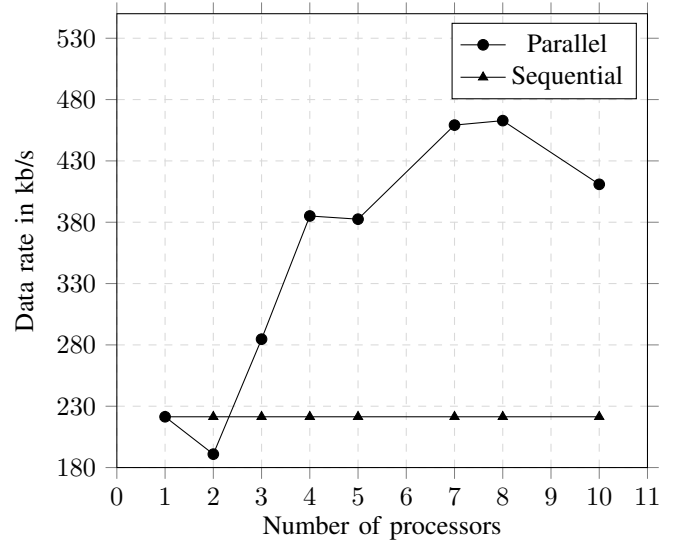
\begin{figure}[tb]
\centering
\begin{tikzpicture}
\begin{axis}[
        width=3.4in, height=3in,     
        grid = major,
        grid style={dashed, gray!30},
        xmin=0,     
        xmax=11,    
        ymin=180,     
        ymax=550,   
        /pgfplots/xtick={0,1,...,11}, 
        /pgfplots/ytick={180,230,...,550}, 
        axis background/.style={fill=white},
        ylabel=Data rate in kb/s,
        xlabel=Number of processors,
        tick align=outside]
      \addplot [mark=*] table [x=nS, y=DR, col sep=comma]{images/MPIdata.csv};
      \addlegendentry{Parallel}
 
      \addplot [mark=triangle*] table [x=nS, y=S, col sep=comma]{images/MPIdata.csv};
      \addlegendentry{Sequential}
    \end{axis}
\end{tikzpicture}
\caption{Throughput rates for sequential and parallel execution of LDPC decoder using MPI framework}
\label{singlescan_throughput_dvb_peggirreg}
\end{figure}

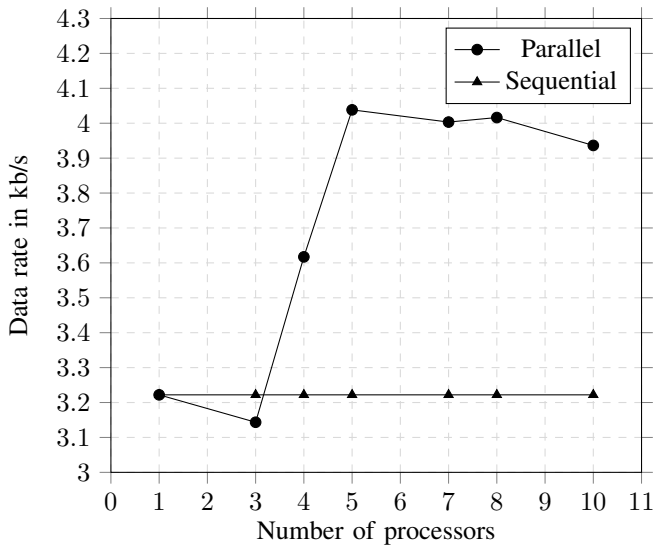
\begin{figure}[tb]
\centering
\begin{tikzpicture}
\begin{axis}[
        width=3.4in, height=3in,     
        grid = major,
        grid style={dashed, gray!30},
        xmin=0,     
        xmax=11,    
        ymin=3,     
        ymax=4.3,   
        /pgfplots/xtick={0,1,...,11}, 
        /pgfplots/ytick={3,3.1,...,4.3}, 
        axis background/.style={fill=white},
        ylabel=Data rate in kb/s,
        xlabel=Number of processors,
        tick align=outside]
      \addplot [mark=*] table [x=nS, y=Par, col sep=comma]{images/HeMPSdata.csv};
      \addlegendentry{Parallel}
 
      \addplot [mark=triangle*] table [x=nS, y=Seq, col sep=comma]{images/HeMPSdata.csv};
      \addlegendentry{Sequential}
    \end{axis}
\end{tikzpicture}
\caption{Throughput rates for sequential and parallel execution of LDPC decoder on HeMPS platform}
\label{singleScanResultsMaxTrial30}
\end{figure}
The results obtained for the MPI framework run on the desktop machine and HeMPS simulations are summarised in Fig.\ref{singlescan_throughput_dvb_peggirreg}, Fig.\ref{singleScanResultsMaxTrial30}, Table \ref{resultsHemps} and
Table \ref{resultsMPI}. The figures and tables give the total number of processors, which includes master processor and slave processors. In our mapping technique, we have chosen to map the check node computations onto slave processors. The master node is in charge of a bulk of processing including variable node computations. The algorithm performs worse initially but the performance increases among the increase of processors compared with the sequential execution. When the number of processors are small, significant processing happens in the slave nodes during which the master has to wait, explaining the initial drop in the performance.

\begin{table}[tb]
\caption{Throughput and speed up factor for simulation of LDPC decoder on HeMPS}\label{resultsHemps} 
\begin{tabulary}{\textwidth}{L|C C C}
Scenario & number of processors & Throughput (kbps) & Speed Up factor\\
  \hline
1 & 1 & 3.222 & -\\
2 & 3 & 3.14 & 0.97\\
3 & 4 & 3.61 & 1.12\\
4 & 5 & 4.03 & 1.25\\
5 & 7 & 4.00 & 1.24\\
6 & 8 & 4.01 & 1.24\\
7 & 10 & 3.93 & 1.22\\
\end{tabulary}

\end{table}

\begin{table}[tb] 
\caption{Throughput and speed up factor for simulation of LDPC decoder using MPI Framework on Desktop}\label{resultsMPI}
\begin{tabulary}{\textwidth}{L|C C C}
Scenario & number of processors & Throughput (kbps) & Speed Up factor\\
\hline
1 & 1 &	221.40 &	-\\
2 & 2 & 190.93 & 0.86\\
3 & 3 & 284.67 & 1.28\\
4 & 4 & 385.10 & 1.73\\
5 & 5 & 382.51 & 1.72\\
6 & 7 & 459.19 & 2.07\\
7 & 8 & 462.83 &	2.09\\
8 & 10 & 410.90 & 1.85\\
\end{tabulary}

\end{table}

Considering the scalability limit, the proposed implementation scales
up to five processors after which the performance starts to degrade.
This decrease in performance can be attributed to the communication
cost between the master and slave processors. The communication cost
that is significant enough to counterbalance the gain from the parallel operations of the check nodes. Unlike the MPI framework, HeMPS is a NoC based architecture where physical distance and network communication costs add up with parallelisation.

From the results, for $252 \times 504$ matrix, we can observe accordance of the
practical measurements of HeMPS with theoretical projections of the
MPI software model. The variation in the speedup values are caused by
the differences between the platforms of measurements: the MPI model is based on a software implementation in a desktop environment thus the results are evaluated in a qualitative approach. However the proposed implementation in HeMPS is suitable for resource-limited embedded MPSoC platforms.

\section{CONCLUSION}
This paper presents a viable solution to implement the decoding of Low Density Parity Check Codes in a MPSoC environment. The simulation of the parallel set was performed and the throughput rates and the speedup factor for RMSA decoding algorithm were given. For a given LDPC code, the performance gain was maximal at five processors after which performance degraded. Throughput rates can be improved by increasing the clock speed and/or memory page size in each core. Parallelism can be introduced in the variable node update stage as well to see the impact of performance. In addition, the redesign of the MPI routine in a non-blocking matter would allow a more efficient job scheduling in the master process which would provide significant performance enhancements.



%
\bibliography{paperLdpcJAGS}
\end{document}